\documentclass[10pt]{article}
\usepackage{latexsym}
\usepackage{amssymb}
\usepackage{amsfonts}
\usepackage{amsmath}
\usepackage{theorem}

\newtheorem{theorem}{Theorem}
\newtheorem{lemma}{Lemma}

\newtheorem{definition}{Definition}
\newtheorem{remark}{Remark}
\newtheorem{proposition}{Proposition}

\begin{document}

\title{Mean field dilute ferromagnet I. High temperature and zero
temperature behavior}

\author{Luca De Sanctis
\footnote{Dipartimenti di Matematica e Psicologia, Universit\`a di Bologna; 
P.zza di Porta San Donato 5, 40126 Bologna, Italy;
{\tt<desanctis@dm.unibo.it>}},
Francesco Guerra
\footnote{Dipartimento di Fisica, Universit\`a La Sapienza, 
INFN, sezione di Roma1,
P.le Aldo Moro 2, 00185 Roma, Italy; 
{\tt<francesco.guerra@roma1.infn.it>}}}

\maketitle

\begin{abstract}
We study the mean field dilute model of a ferromagnet.
We find and prove an expression for the free energy density
at high temperature, and at temperature zero. We find the critical
line of the model, separating the phase with zero magnetization
from the phase with symmetry breaking. We also compute exactly
the entropy at temperature zero, which is strictly positive. 
The physical behavior at temperature zero is very interesting 
and related to infinite dimensional
percolation, and suggests possible behaviors at generic low temperatures.
Lastly, we provide a complete solution for the annealed model.
Our results hold both for the Poisson and the Bernoulli versions
of the model.
\end{abstract}



\section{Introduction}

The study of the Curie-Weiss mean field ferromagnetic model
can be performed by very simple convexity methods \cite{g7, g8}.
The main ingredient of such approaches is the quadratic
dependence of the Hamiltonian on the magnetization, which
makes the free energy convex in such a quantity. A very strong
analogy allows to reproduce the same physical approach 
when the ferromagnet interaction is replaced
by one with Gaussian distribution
\cite{g7, g8}. In this case the convexity arguments apply through
the quadratic dependence of the \emph{covariance} 
of the Hamiltonian on the
main physical quantity for these models: the overlap between two 
configurations. Despite the initial belief that the general approach
reviewed in \cite{g7, g8} was possible only because of the 
special nature of Gaussian interactions, it turned out
that the mentioned analogy extends to dilute mean field
spin glasses \cite{fl, lds1, lds5}, and independetly of the (symmetric)
distribution of the interactions.
Surprisingly, the case of
dilute mean field ferromagnets has not been framed so far
in the context of these methods, reviewed in \cite{g7, g8, lds5}.
This is not the only reason why the mean field dilute ferromagnet
is an interesting model, as we will see. 
It is also noticeable 
that the mean field dilute ferromagnet 
has not been studied so much.
Most of the studies in the physical literature 
are not fully rigorous and only graze the dilute ferromagnet 
within works about different models, and often within a general study
of models on various kinds of networks, which gained recently a large 
attention (see for instance \cite{dgm, ks, has} and references therein).
An exception is \cite{gm}, which is a rigorous quite general study 
about reconstruction for models on random graphs, but also analyzes the
case of an Ising ferromagnet. Our methods and purposes 
are different from those of \cite{gm}, where 
results of quite general nature are present about the existence of limits.
A clear explanation of the physical behavior
of the dilute mean field ferromagnet is still missing.

We started our work studying the annealed
version of the mean field dilute ferromagnet, which is
simpler than the quenched one and it
can be studied with convexity techniques which generalize
those illustrated in \cite{g7, g8}. This study is reported in an appendix.
Despite its simplicity, the annealed
model exhibits a rich behavior. We then studied rigorously
the quenched model using quite simple techniques, revealing
interesting connections with glassy systems. Our model is
therefore a long awaited one with intermediate difficulty and behavior
between fully connected ferromagnets and spin glasses. We prove 
that there is a high temperature region with identically zero magnetization,
delimited by a critical line which we find, in which we can
compute the free energy exactly. We also find and prove
an exact expression for the free energy and for the entropy at temperature
zero. Our approach suggests possible behaviors and techniques
to deal with generic low temperatures, on which we will report soon \cite{dsg2}. 
Lastly, we briefly show some self-averaging properties of the model,
which are of general interest in thermodynamics, and in the case of the
magnetization are used here to control the model at temperature zero.

We focus on the Poisson version of the model, but the main results
hold in the Bernoulli version as well (see \cite{sv} for more details).


\section{The model and some preliminary facts}

In this section we introduce the model and the notations,
and provide some useful formulas which are at the basis of
almost all the calculations needed in this article.

\subsection{Definitions}

Given a set of $N$ points, the model is defined through
configurations $\sigma: i\to \pm 1$, $i=1,\dots , N$
of Ising spins. By $\{i_{\nu},j_{\nu},k_{\nu},l_{\nu}\}$ we will
denote families independent random variables all uniformly
distributed on $1,\ldots , N$.
The Hamiltonian is the random function defined by
$$
H_{N}(\sigma)=-\sum_{\nu=1}^{K}\sigma_{i_{\nu}}\sigma_{j_{\nu}}
$$
where $K$ is a Poisson random variable of mean $\alpha N$,
for some given $\alpha\in\mathbb{R}_{+}$, which is called
\emph{connectivity}.
The expectation with respect to the random choice of the spins
and with respect
to the Poisson random variable is denoted by $\mathbb{E}$,
and it is called \emph{quenched} expectation.
Given a non-negative real number $\beta$, the function
$$
A_{N}(\alpha,\beta)=\frac1N\mathbb{E}\ln\sum_{\sigma}
\exp(-\beta H_{N}(\sigma))
$$
is called pressure, and $-A_{N}(\alpha,\beta)/\beta$ is the free energy.
Given the simple relation between the two, we will indifferently use either
one or the other.
The sum
$$
Z_{N}(\beta)=\sum_{\sigma}
\exp(-\beta H_{N}(\sigma))
$$
is the (random) partition function, and the Boltzmann-Gibbs 
expectation of an observable $\mathcal{O}:\sigma\to \mathcal{O}(\sigma)\in\mathbb{R}$ is
$$
\Omega(\mathcal{O})=\frac{1}{Z_{N}}\sum_\sigma \exp(-\beta H_{N}(\sigma))
\mathcal{O}(\sigma)\ .
$$
When it is not confusing, we will omit the dependence
of $\Omega$ on $N$ or on the Poisson random variable appearing
in the Hamiltonian. When we omit the index $N$ in the pressure
we mean to have taken the thermodynamic limit.
The main physical quantity in this model is the magnetization of a configuration
$$
m(\sigma)=\frac1N \sum_{i=1}^{N}\sigma_{i}\ .
$$
A further notation is $\langle \cdot \rangle = \mathbb{E}\Omega(\cdot)$.
Throughout the paper, $t\in[0,1]$ will be a real interpolating parameter,
and $\delta_{AB}$ is the Kroneker function, equal to one, if $A=B$, equal
to zero otherwise.

A simple calculation immediately provides the following
useful form for the pressure
\begin{equation}
\label{alternativa}
A_{N}(\alpha,\beta)=\alpha\ln\cosh\beta+\frac1N\mathbb{E}\ln
\sum_{\sigma}\prod_{\nu=1}^{K}
(1+\sigma_{i_{\nu}}\sigma_{j_{\nu}}\tanh\beta)\ .
\end{equation}

Notice that the ferromagnetism implies that the pressure of the Poisson
model is always larger than or equal to the one of the Bernoulli version.
Notice also that not much changes in the Poisson model if 
one considered a sort of truncated distribution, 
in which the probability of the integer one is left
unchanged and the only other possible value is zero. This preserves the main
feature of the formula at the basis of the next subsection, and allows
to compare this truncated model with the Bernoulli one, whose pressure
is smaller. This is therefore a way to compare the Bernoulli model with the 
Poisson one (more general considerations can be found in \cite{sv}).

\subsection{Properties of the Poisson measure}

Given a function $g:\mathbb{N}\to\mathbb{R}$ 
and a generic Poisson variable $K$ with mean $\zeta$,
whose expectation is denoted by $\mathbb{E}$,
it is easy to verify that
\begin{equation}
\label{main1}
\mathbb{E}[Kg(K)]=\zeta\mathbb{E}[g(K-1)]
\end{equation}
and that
\begin{equation}
\label{main2}
\frac{d}{d\zeta}\mathbb{E}[g(K)]=\mathbb{E}[g(K+1)-g(K)]\ .
\end{equation}
Along the same lines it is interesting to note that
the second derivative resembles a Laplacian
\begin{equation}
\label{laplacian}
\frac{d^{2}}{d\zeta^{2}}\mathbb{E}g(K)=\mathbb{E}[g(K+2)-2g(K+1)+g(K)]\ .
\end{equation}

These formulas are used very often in the calculations we need in 
the present work.

\subsection{The main derivatives}

A simple use of (\ref{main1}) yields
\begin{equation}
\label{betaderivative}
\frac{\partial A_{N}(\alpha,\beta)}{\partial\beta}=
\alpha\mathbb{E}\frac{\tanh\beta+
\Omega(\sigma_{i_{0}}\sigma_{j_{0}})}{1+
\Omega(\sigma_{i_{0}}\sigma_{j_{0}})\tanh\beta}\ ;
\end{equation}
where the index $0$ is henceforth used
to denote random variables independent of those
appearing in $\Omega$.
A simple use of (\ref{main2}) yields instead
\begin{equation}
\label{alphaderivative}
\frac{\partial A_{N}(\alpha,\beta)}{\partial\alpha}=\ln\cosh\beta+
\mathbb{E}\ln[1+\Omega(\sigma_{i_{0}}\sigma_{j_{0}})\tanh\beta]\ .
\end{equation}
These two derivatives will be constantly used in the present work.


\section{Some basic properties of the model}

\subsection{Equivalent formulation}

We want to show that the Hamiltonian can be written
in three different forms, equivalent in distribution:
$$
-H_{N}(\sigma)=\sum_{\nu=1}^{K}\sigma_{i_{\nu}}\sigma_{j_{\nu}}
\sim \sum_{i=1}^{N}\left(\sum_{\nu=1}^{K^{i}}\sigma_{j_{\nu}}\right)\sigma_{i}
\sim \sum_{i,j}^{1,N}\sum_{\nu=1}^{K^{ij}}\sigma_{i}\sigma_{j}
= \sum_{i,j}^{1,N}K^{ij}\sigma_{i_{\nu}}\sigma_{j_{\nu}}
$$
where $K$ is a Poisson random variable of mean $\alpha N$,
$\{K^{i}\}$ are independent Poisson random variables with mean
$\alpha$, $\{K^{ij}\}$ are independent Poisson random variables
with mean $\alpha /N$.
Let us explain what we mean by ``equivalent'' from the point of view of 
the thermodynamics of our model. We will do so by means of interpolation.
We will henceforth use the same $\Omega$ for the Gibbs measure, even when
the Boltzmannfaktor is not associated with the original Hamiltonian of the model,
but with any generic Hamiltonian (for instance an interpolating one). The weights
defining the Gibbs measure are each time easily deducible from the calculations.
Take
\begin{equation*}
\Phi_{N}(t)=\frac1N\mathbb{E}\ln\sum_{\sigma}\exp
\beta[\sum_{\nu=1}^{K_{1}}\sigma_{i_{\nu}}\sigma_{j_{\nu}}
+\sum_{i,j}^{1,N}K_{0}^{ij}\sigma_{i}\sigma_{j}]\ ,
\end{equation*}
where $K_{1}$ is a Poisson random variable with mean $t\alpha N$, 
and $\{K_{0}^{ij}\}$ are independent Poisson random  variables with mean
$(1-t)\alpha / N$. 
Compute now the derivative with respect to $t$
$$
\frac{1}{\alpha}\frac{d\Phi_{N}(t)}{dt}=
\mathbb{E}\ln(1+\Omega(\sigma_{i_{0}}\sigma_{j_{0}})\tanh\beta)
-\frac{1}{N^{2}}\sum_{i,j}^{1,N}\ln(1+\Omega(\sigma_{i}\sigma_{j})\tanh\beta)=0
$$
where the equality of the two terms is due to the expectation with
respect to $i_{0}$ and $j_{0}$, which is included in the expectation $\mathbb{E}$
with respect to all the quenched random variables (recall that
$i_{0},j_{0}$ are independent of the random site indices in the $t$-dependent
$\Omega$).



\subsection{Convexity of the pressure as a function of the connectivity}

In this subsection we want prove the following
\begin{proposition}
The pressure $A_{N}(\alpha, \beta)$ is a convex function of the
connectivity $\alpha$.
\end{proposition}
Here
it is useful to specify on which Poisson random variable the 
Gibbs measure depends on, we will do so by means of an index.
By $K$ we denote as usual a Poisson random variable of mean
$\alpha N$.\\
\textbf{Proof}. 
Employing (\ref{laplacian}) one finds
\begin{multline*}
\frac{\partial^{2}A_{N}(\alpha,\beta)}{\partial\alpha^{2}}=
\frac{\partial}{\partial\alpha}\mathbb{E}\ln[1+\Omega_{K}(\sigma_{i_{0}}
\sigma_{j_{0}})\tanh\beta]\\
=N\mathbb{E}\ln\left[\frac{1+\Omega_{K+1}(\sigma_{i_{0}}
\sigma_{j_{0}})
\tanh\beta}{1+\Omega_{K}(\sigma_{i_{0}}\sigma_{j_{0}})\tanh\beta}\right]
\end{multline*}
where $K$ is the usual Poisson random variable with mean $\alpha N$.
It is also easy to check that
$$
\Omega_{K+1}(\sigma_{i_{0}}\sigma_{j_{0}})=
\frac{\Omega_{K}(\sigma_{i_{0}}\sigma_{j_{0}})+
\Omega_{K}(\sigma_{i_{0}}\sigma_{j_{0}}\sigma_{k_{0}}
\sigma_{l_{0}})\tanh\beta}{1+\Omega_{K}(\sigma_{k_{0}}\sigma_{l_{0}})
\tanh\beta}\ .
$$
From these last two expressions, after a few calculations,
one obtains
\begin{multline*}
\frac{1+\Omega_{K+1}(\sigma_{i_{0}}
\sigma_{j_{0}})
\tanh\beta}{1+\Omega_{K}(\sigma_{i_{0}}\sigma_{j_{0}})\tanh\beta}=\\
1+(\tanh^{2}\beta)
\frac{\Omega(\sigma_{i_{0}}\sigma_{j_{0}}
\sigma_{k_{0}}\sigma_{l_{0}})-\Omega(\sigma_{i_{0}}\sigma_{j_{0}})
\Omega(\sigma_{k_{0}}\sigma_{l_{0}})}{(1+
(\tanh\beta)\Omega(\sigma_{i_{0}}\sigma_{j_{0}}))(1+(\tanh\beta)
\Omega(\sigma_{k_{0}}\sigma_{l_{0}}))}\ .
\end{multline*}
Hence
$$
\frac{\partial^{2}A_{N}(\alpha,\beta)}{\partial\alpha^{2}}\geq 0
$$
due to the ferromagnetic nature of the interactions, which implies 
$$
\Omega(\sigma_{i_{0}}\sigma_{j_{0}}
\sigma_{k_{0}}\sigma_{l_{0}})-\Omega(\sigma_{i_{0}}\sigma_{j_{0}})
\Omega(\sigma_{k_{0}}\sigma_{l_{0}})\geq 0
$$
and completes the proof. $\Box$

\subsection{The infinite connectivity limit}

Recall that the Hamiltonian of the Curie-Weiss (CW) model
is 
$$
H^{(CW)}_{N}(\sigma)=\frac12 Nm^{2}(\sigma)\ ,
$$ 
and the associated pressure will be denoted by
$A_{N}^{(CW)}(\beta)$.
Given $n$ configurations $\sigma^{(1)},\ldots ,\sigma^{(n)}$,
let us define the multi-overlaps by
$$
q_{n}=\frac1N\sum_{i=1}^{N}\sigma_{i}^{1}\cdots\sigma_{i}^{(n)}\ .
$$
Notice that $q_{1}=m$ is the magnetization.
Let us prove the next
\begin{proposition}
If we let $\alpha\to\infty$, $\beta\to 0$ 
with $2\alpha\tanh\beta=\beta^{\prime}$
kept constant, the pressure $A_{N}(\alpha,\beta)$
tends to the one of the CW model:
$$
\lim_{\begin{array}{c}
 \vspace{-2mm}  \scriptscriptstyle\vspace{-2mm} \alpha\to\infty   \\
  \vspace{-2mm} \scriptscriptstyle\vspace{-2mm} \beta\to\infty  \\
   \scriptscriptstyle 2\alpha\tanh\beta=\beta^{\prime}
\end{array}}A_{N}(\alpha,\beta)=A^{(CW)}_{N}(\beta^{\prime})
$$
uniformly in the size of the system.
\end{proposition}
A more general theorem has been proven long time ago in \cite{bg}.\\
\textbf{Proof}. Consider the following interpolation
\begin{equation}
\label{interpolazione}
\hat{A}_{N}(t)\equiv\frac1N\mathbb{E}\ln\sum_{\sigma}\exp\bigg[
\beta\sum_{\nu=1}^{K_{1}}\sigma_{i_{\nu}}\sigma_{j_{\nu}}
+(1-t)\beta^{\prime} \frac12 Nm^{2}\bigg]
\end{equation}
between the Curie-Weiss model and its dilute version, 
$K_{1}$ being a Poisson random variable with mean $t\alpha N$.
Clearly $\hat{A}_{N}(0)=A_{N}^{(CW)}(\beta^{\prime})$
and $\hat{A}_{N}(1)=A_{N}(\alpha,\beta)$.
A series expansion of (\ref{alphaderivative}) offers
\begin{equation}
\label{paragone}
\frac{d\hat{A}_{N}(t)}{dt}=
\alpha\ln\cosh\beta-\alpha\sum_{n}\frac{(-1)^{n}}{n}
\langle q^{2}_{n}\rangle\tanh^{n}\beta-
\frac12\beta^{\prime} \langle m^{2} \rangle\ ,
\end{equation}
where the average $\langle\cdots\rangle$ depends of $t$
because of the interpolation.
When
$\alpha\to\infty$, $\beta\to 0$ with $2\alpha\tanh\beta=\beta^{\prime}$
all the $\alpha\tanh^{n}\beta\to 0$ for $n>1$, and
the right hand side of (\ref{paragone}) reduces to 
$\alpha\tanh\beta\langle m^{2} \rangle-\beta^{\prime}\langle m^{2} \rangle/2
= 0$. $\Box$ \\
Since at the order $n=1$
the quenched and annealed model coincide, as deducible from the
previous subsection (see the appendix), the same identical proof holds for the
annealed model as well, where $\Omega(\sigma_{i_{0}}\sigma_{j_{0}})$
is replaced by $\Omega(m^{2})$.
\begin{remark}
The dilute model, be it quenched or annealed, 
reduces to the fully connected one in the
infinite connectivity limit uniformly in the size of the system.
\end{remark}

\subsection{The ground state}

From (\ref{betaderivative})
and 
$$
\frac{\partial A_{N}(\alpha,\beta)}{\partial\beta}=-\frac{\langle H_{N}\rangle}{N}\ ,
$$
we immediately get
$$
\lim_{\beta\to\infty}\frac{\langle H_{N}\rangle}{N}=-\alpha
$$
for all $N$, and clearly
$$
\lim_{\beta\to\infty}\frac{1}{\beta}\frac{\partial A_{N}(\alpha,\beta)}{\partial\beta}=0\ .
$$
The same ground state is well reproduced by the annealed model (see the appendix),
since these formulas stay the same.


\section{High temperature, low connectivity, symmetric region}

In this section we show that at least in a region with small
enough connectivity we can compute the free energy in the
thermodynamic limit. We do not prove here the existence of any
symmetry breaking nor that the considered region
spans the whole symmetric phase. These aspects will be studied 
later on, here we are about to prove only the next
\begin{theorem} In the region defined by
$$
2\alpha\tanh\beta\leq 1
$$
the thermodynamic limit of the pressure is given by
$$
A(\alpha,\beta)=\ln 2 +\alpha\ln\cosh\beta\ .
$$
In particular, when $\alpha\leq 1/2$ any value of the inverse temperature
$\beta$ fulfills the condition.
\end{theorem}
\textbf{Proof}.
First of all notice that (\ref{alphaderivative}) implies
$$
\frac{\partial A_{N}(\alpha,\beta)}{\partial\alpha}\geq \ln\cosh\beta
$$
since the ferromagnetism imposes
$\Omega(\sigma_{i_{0}}\sigma_{j_{0}})\geq 0$.
As a consequence, we get immediately a first inequality
$$
A_{N}(\alpha,\beta)\geq \ln 2+\alpha\ln\cosh\beta\ .
$$
Now consider again the interpolation (\ref{interpolazione})
and use (\ref{alternativa}) to observe that
\begin{eqnarray*}
\frac{d\hat{A}_{N}(t)}{dt} & = &  
\alpha[\ln\cosh\beta+\mathbb{E}\ln(1+
\Omega(\sigma_{i_{0}}\sigma_{j_{0}})\tanh\beta]-
\beta^{\prime}\mathbb{E}\frac12\Omega(m^{2})\\
{} & \leq & \alpha\ln\cosh\beta+\alpha\mathbb{E}
\Omega(\sigma_{i_{0}}\sigma_{j_{0}})\tanh\beta-
\beta^{\prime}\mathbb{E}\frac12\Omega(m^{2})\\
{} & = & \alpha\ln\cosh\beta
\end{eqnarray*}
since $\ln(1+x)\leq x$ and $2\alpha\tanh\beta=\beta^{\prime}$. 
Therefore, by the fundamental theorem of calculus
\begin{equation}
\label{rottura}
A_{N}(\alpha,\beta)\leq\alpha\ln\cosh\beta+A_{N}^{(CW)}(\beta^{\prime})\ .
\end{equation}
Now, it is well known \cite{g7,g8} that
$$
\lim_{N\to\infty}A_{N}^{(CW)}(\beta^{\prime})= \ln 2\ \Leftarrow\ \beta^{\prime}\leq 1
$$
and therefore putting together the two opposite inequalities we obtain
$$
2\alpha\tanh\beta\leq 1\ \Rightarrow\ 
\lim_{N\to\infty}A_{N}(\alpha,\beta)=\ln 2 +\alpha\ln\cosh\beta
$$
and the proof is complete. $\Box$

\begin{remark}
We obtained \emph{en passant} the existence of the
thermodynamic limit of the free energy per spin at least in the
considered region.
\end{remark}
\begin{remark}
The pressure of the Viana-Bray model at high temperature
is the same, such an expression being of quite general validity.
\end{remark}

This low connectivity behavior
is well described by the annealed approximation, discussed in the appendix.


\section{The model at temperature zero}

We have just seen that in a high temperature region the 
free energy can be easily computed explicitly. We will now study
the model at temperature zero. This will also be used to prove
that the described high temperature region cannot be extended, 
in the sense that outside such a region the solution
we obtained above does not hold. 

Let us introduce the following notation
for the difference between the pressure and its
high temperature expression:
\begin{eqnarray*}
\label{}
\tilde{A}_{N}(\alpha,\beta) & = & A_{N}(\alpha,\beta)-\ln 2-\alpha\ln\cosh\beta \\
{} & = & \frac1N\mathbb{E}\ln\sum_{\sigma}
\prod_{i,j}^{1,N}(1+\sigma_{i}\sigma_{j}\tanh\beta)^{K^{ij}}-\ln 2\ .
\end{eqnarray*}
We have just seen in the previous subsection that in the thermodynamic limit
$\tilde{A}_{N}(\alpha,\beta)\to 0$ if the temperature is such that
$2\alpha\tanh\beta\leq 1$. We will see in the next section that
this quantity is different from zero outside the region. In this section
we want to study the model at temperature zero, and we will need
$$
\tilde{A}_{N}^{(0)}(\alpha)=
\liminf_{\beta\to\infty}\tilde{A}_{N}(\alpha,\beta)
=\frac1N\mathbb{E}\ln\sum_{\sigma}
\prod_{i,j}^{1,N}(1+\sigma_{i}\sigma_{j})^{K^{ij}}-\ln 2  
$$
Let us show that the thermodynamic limit of this quantity
exists. This is guaranteed by the next
\begin{lemma}\label{existence}
The function $N\tilde{A}_{N}^{(0)}(\alpha)$ 
is sub-additive in the size  of the system $N$.
\end{lemma}
\textbf{Proof}. The proof can be obtained through interpolation,
and it is guided by the reasonings described in the next subsection.
Decompose the system into two subsystems
of sizes $N_{1}$ and $N_{2}$, and denote by
$\sigma^{\prime}$ the spins in the first block,
by $\sigma^{\prime\prime}$ the spins of the second
block, while $\sigma$ will still denote configurations
of the whole system.
Define, for $t\in[0,1]$,
$$
\psi(t)=\mathbb{E}\ln\sum_{\sigma}
\prod_{i,j}^{1,N}(1+\sigma_{i}\sigma_{j})^{K_{0}^{ij}}
\prod_{i,j}^{1,N}(1+\sigma^{\prime}_{i}\sigma^{\prime}_{j})^{K_{1}^{ij}}
\prod_{i,j}^{1,N}(1+\sigma^{\prime\prime}_{i}\sigma^{\prime\prime}_{j})^{K_{2}^{ij}}
$$
where $\{K_{0}^{ij}\}$, $\{K_{1}^{ij}\}$, $\{K_{2}^{ij}\}$, are families of independent
Poisson random variables with mean $t\alpha/N$, $(1-t)\alpha/N_{1}$, 
$(1-t)\alpha/N_{2}$ respectively. Denote by $m_{1}$ and $m_{2}$
the magnetizations of the two blocks.
A direct calculation gives
$$
\frac{d}{dt}\psi(t)=N\ln 2\bigg[\mathbb{E}\Omega(m^{2})
-\frac{N_{1}}{N}\mathbb{E}\Omega(m_{1}^{2})
-\frac{N_{2}}{N}\mathbb{E}\Omega(m_{2}^{2})\bigg]\leq 0
$$
by convexity (the next subsection explains the mechanism at the basis
of this result). This allows a comparison
between the values of $\psi(t)$ at zero and at one, which describe
the dependence of $\tilde{A}_{N}^{(0)}$ on the volume involved in 
the sub-additivity. So the proof is complete. $\Box$

The convexity just seen will let us compute the entropy and
the free energy explicitly.

\subsection{Free energy}

Consider again 
$$ 
\tilde{A}_{N}^{(0)}(\alpha)=
\frac1N\mathbb{E}\ln\sum_{\sigma}
\prod_{i,j}^{1,N}(1+\sigma_{i}\sigma_{j})^{K^{ij}}-\ln 2 \ .
$$
It is obvious that if some $K^{ij}\geq 1$ then only the configurations
with $\sigma_{i}=\sigma_{j}$ contribute. In fact the only alternative would
be $\sigma_{i}=-\sigma_{j}$ implying $(1+\sigma_{i}\sigma_{j})^{K^{ij}}=0$.
Therefore for a given realization of the $\{K^{ij}\}$ the set of spins
decomposes into, say, $L$ non-empty clusters of sizes $N_{1},\ldots , N_{L}$,
such that all the spins in a given cluster share the sign and are connected
by non-zero links $K^{ij}$. The quantity $\Omega(\sigma_{i}\sigma_{j})$
can thus take only the values zero and one, according to whether the sites
$i$ and $j$ are connected or not (i.e. belong to the same cluster or not
or equivalently $K^{ij}$ is different or equal to zero).
In the fully connected Curie-Weiss model there is only one cluster coinciding
with the whole system of $N$ spins, all aligned.
The fact that $\Omega(\sigma_{i}\sigma_{j})$ is either zero or one implies
$$
\frac{d}{d\alpha}\tilde{A}^{(0)}_{N}(\alpha)=\mathbb{E}\ln(1
+\Omega(\sigma_{i}\sigma_{j}))=\mathbb{E}\Omega(\sigma_{i}\sigma_{j})\ln 2
=\mathbb{E}\Omega(m^{2})\ln 2
$$
and
$$
\mathbb{E}\Omega(m^{2})=\frac{1}{N^{2}}\mathbb{E}(N_{1}^{2}+\cdots
+N_{L}^{2})\ .
$$
We can also write
$$
\tilde{A}^{(0)}_{N}(\alpha)=\frac1N\mathbb{E}\ln\sum_{\sigma}
\prod_{i,j}^{1,N}\left(\frac{1+\sigma_{i}\sigma_{j}}{2}\right)^{K^{ij}}
-(1-\alpha)\ln 2
$$
and notice
$$
\left(\frac{1+\sigma_{i}\sigma_{j}}{2}\right)^{K^{ij}}=\bigg\{
\begin{array}{ll}
     1\ , & \mbox{if}\ \  K^{ij}=0\ ,  \\
     \delta_{\sigma_{i}\sigma_{j}}\ , & \mbox{if}\ \  K^{ij} > 0  \ .
\end{array}
$$
Hence all strictly positive values of the Poisson variables yield
the same identical contribution, and therefore at least at temperature
zero it is trivial to see that our Poisson model is equivalent
 to a Bernoulli one, in which
the couplings obey
$\bar{K}^{ij}=0$ with probability $\bar{p}_{0}=\exp(-\alpha/N)$, and
$\bar{K}^{ij}=1$ with probability $\bar{p}_{1}=1-\bar{p}_{0}=\sum_{k>0}p_{k}$.
The notation we just used distinguishes the Bernoulli 
case from the Poisson one by means of the bar,
both for the weights $\bar{p}_{k},p_{k}$ and for the random variables 
$\bar{K}^{ij}, K^{ij}$; $k$ is clearly a natural number, and
$p_{k}=\exp(-\alpha/N)\alpha^{k}/(N^{k}k!)$.
For large $N$ only the dominant terms contribute, and we could
equivalently take $\bar{p}_{0}=1-\alpha/N$, $\bar{p}_{1}=\alpha/N$.
Summarizing:
\begin{remark}
The results we obtain regarding the entropy at temperature zero 
hold both in the case of Bernoulli dilution and in the Poisson one.
For statements of wider validity see \cite{sv}.
\end{remark}
Notice that at any temperature the ferromagnetism implies that
the Poisson model  
gives an upper bound for the Bernoulli one, since 
larger values of $K^{ij}$ (possible only in the Poisson case) 
increase the pressure. The opposite bound proving the equivalence
between the two versions of the model is more involved, and relies
on the fact that the mean of the coupling variables $K^{ij}$
is proportional to $1/N$, so that values larger than one 
tend to be negligible events and the remaining two possibilities
give the same contribution in the Poisson and Bernoulli cases.
\begin{remark}
The decomposition into clusters of the spins does not depend on the 
temperature, being determined by the random couplings only.

Moreover, it turns out that the clusters into which the systems decomposes
are dominated by a very large one, surrounded by many small ones,
and this is connected with infinite dimensional percolation.
\end{remark}
Notice that the magnetization of the $l$-th cluster is $N_{l}/N$.

The main purpose of this subsection is to prove the next
\begin{theorem}\label{zeropressure}
At temperature zero, the pressure per spin of the dilute mean field
ferromagnet, be it Poisson or Bernoulli, is given by the formula
$$
\lim_{N\to\infty}\lim_{\beta\to\infty}\frac1N\mathbb{E}\ln Z_{N}(\beta)=\max_{M}
\{2\alpha M +2\alpha\exp(-2\alpha M)-\alpha M^{2}\}\ln 2
$$
in the thermodynamic limit, where $\alpha$ is the degree
of connectivity of the system.
\end{theorem}
The value of $M$ where the maximum is attained is such that
$$
M=1-\exp(-2\alpha M)\ ,
$$
and exhibits a critical value $\alpha=1/2$, below which 
$M$ is equal to zero, above it is different from zero. We will
get back to this along the proof.

The statement and proof of this theorem provide a connection
between statistical mechanics and graph theory \cite{jlr}.\\
\textbf{Proof}. The theorem will be proven through two opposite
bounds.
The convexity we found at the basis of the sub-additivity of 
$N\tilde{A}_{N}^{(0)}(\alpha)$ makes it possible to introduce
a ``replica symmetric cavity'', whose bound will turn out to 
be exact. \\
\textbf{First bound}.
Define, for $t\in[0,1]$
$$
\varphi_{N}(t)=\frac1N \mathbb{E}\ln\sum_{\sigma}\prod_{ij}^{1,N}
(1+\sigma_{i}\sigma_{j})^{K^{ij}_{1}}\prod_{i=1}^{N}(1+\sigma_{i})^{K^{i}_{0}}
$$
where $K^{ij}_{1}$ are independent Poisson random variables
with mean $t\alpha/N$, $K^{i}_{0}$ are independent Poisson random
variables with mean $(1-t)2M\alpha$ with $M$ a free parameter.
If $K^{i}_{0}=0$, then $(1+\sigma_{i})^{K^{i}_{0}}=1$. If instead 
$K^{i}_{0}>0$, then $(1+\sigma_{i})^{K^{i}_{0}}=\delta_{1\sigma_{i}}2^{K^{i}_{0}}$.
Which means the spin $\sigma_{i}$ is forced to take the value one.
Hence the system decomposes into blocks with 
non-zero internal links $K^{ij}_{1}$, and within each of these blocks
all the spins are equal to one if $K^{i}_{0}=0$ all values of $i$,
while the spins are free to fluctuate if $K^{i}_{0}>0$ for at least
one value of $i$. Notice that $\Omega(\sigma_{i}\sigma_{j})=0,1$
and $\Omega(\sigma_{i})=0,1$, moreover for $t=0$ the function
$\varphi_{N}$ can be computed explicitly.
Let us calculate its derivative
\begin{eqnarray}
\frac{d}{dt}\varphi_{N}(t) & = & \alpha\mathbb{E}\ln(1+\Omega_{t}[\sigma_{i}])
-2\alpha M\mathbb{E}\ln(1+\Omega_{t}[\sigma_{i}]) \nonumber\\
{} & = & \alpha \ln 2 (\mathbb{E}\Omega_{t}[m^{2}]-2M\mathbb{E}\Omega_{t}[m])
\nonumber\\
{} & = &  \alpha \ln 2 (\mathbb{E}\Omega_{t}[(m-M)^{2}])-\alpha M^{2} \ln 2
\label{somma}
\end{eqnarray}
with an obvious meaning of $\Omega_{t}$. Hence
$$
\frac{d}{dt}\varphi_{N}(t)\geq -\alpha M^{2}\ln 2\ .
$$
Integrating between zero and one
$$
\varphi_{N}(1)=\frac1N \mathbb{E}\ln Z_{N}\geq \varphi_{N}(0)
-\alpha M^{2}\ln 2\ .
$$
The computation of $\varphi_{N}(0)$ is not difficult, if we notice
that
\begin{eqnarray*}
&& \sum_{\sigma_{i}}(1+\sigma_{i})^{K^{i}_{0}}=2 \qquad \mbox{if} \quad K^{i}_{0}=0 \\
&& \sum_{\sigma_{i}}(1+\sigma_{i})^{K^{i}_{0}}=2^{K^{i}_{0}} \quad \mbox{if}   
\quad K^{i}_{0}>0
\end{eqnarray*}
Therefore
$$
\varphi_{N}(0)=\mathbb{E}\ln\sum_{\sigma}(1+\sigma_{i})^{K^{i}_{0}}=
\tilde{p}_{0}\ln 2 + \sum_{k=1}^{\infty}\tilde{p}_{k}k\ln 2=
[\exp(-2\alpha M)+2\alpha M]\ln2
$$
where $\tilde{p}_{0}=\exp(-2\alpha M)$ and in the sum over the integer $k$
the term corresponding to $k=0$ can be added.
Hence
$$
\lim_{\beta\to\infty}\frac1N\mathbb{E}\ln Z_{N}(\beta)\geq 
[2\alpha M+\exp(-2\alpha M)-\alpha M^{2}]\ln 2\equiv [\tilde{\varphi}(M)]\ln 2\ ,
$$
which proves the first bound for any size $N$ of the system.
Notice that the derivative with respect to $M$ of $\tilde{\varphi}(M)$ is
$$
\tilde{\varphi}^{\prime}(M)=2\alpha-2\alpha\exp(-2\alpha M)-2\alpha M
=2\alpha M\left(\frac{1-\exp(-2\alpha M)}{M}-1\right)\ .
$$
The function $(1-\exp(-2\alpha M))/M$ is decreasing in $M$ and so is
$d\tilde{\varphi}(M)/dM^{2}$, and $\tilde{\varphi}(M)$ is a concave
function of $M^{2}$. Its maximum is at zero if $\alpha\leq 1/2$, it is
different from zero if $\alpha>1/2$, and more precisely where
\begin{equation}
\label{extremal}
1-M=\exp(-2\alpha M)\ .
\end{equation}
Let us recap what we proved
$$
\lim_{\beta\to\infty}\lim_{N\to\infty}A_{N}(\alpha,\beta)\geq 
\sup_{M}\{2\alpha M+\exp(-2\alpha M)-\alpha M^{2}\}\ln 2\ .
$$
We are now going to show that the opposite bound holds 
in the thermodynamic limit.\\
\textbf{Second bound}. We will make use of the self-averaging of the
magnetization, proven in subsection \ref{self-av}.
Let start from the sum rule
\begin{multline*}
\frac1N \mathbb{E}\ln\sum_{\sigma}(1+\sigma_{i}\sigma_{j})^{K^{ij}}=\\
[2\alpha M+\exp(-2\alpha M)-\alpha M^{2}]\ln2+\alpha\ln 2\int_{0}^{1}
\mathbb{E}\Omega_{t}[(m-M^{2})]dt
\end{multline*}
which is a consequence of (\ref{somma}) and
the expression of $\varphi_{N}(0)$ we computed.
We want to show that 
$$
\lim_{n\to\infty}\int_{0}^{1}\mathbb{E}\Omega_{t}[(m-M)^{2}]dt = 0 \ .
$$
Given a statement $I$, let us define the truth function
$\chi_{I}$ as equal to one if $I$ is true, equal to zero if $I$ is false.
Let us then split the integral into three pieces:
\begin{eqnarray*}
\int_{0}^{1}\mathbb{E}\Omega_{t}[(m-M)^{2}]dt & = &  
\int_{0}^{1}\mathbb{E}\Omega_{t}[(m-M)^{2}\chi_{m\leq M-\epsilon}] dt\\
{} & {} & +\int_{0}^{1}\mathbb{E}
\Omega_{t}[(m-M)^{2}\chi_{M-\epsilon\leq m\leq M+\epsilon}] dt \\
{} & {} & + \int_{0}^{1}\mathbb{E}
\Omega_{t}[(m-M)^{2}\chi_{m\geq M+\epsilon}] dt\ .
\end{eqnarray*}
The second of the three terms in the right hand side is clearly bounded by
$\epsilon^{2}$. We want to show that the other two terms vanish
in the thermodynamic limit for any $\epsilon >0$, so to have
$$
\lim_{N\to\infty}A_{N}(\alpha,\infty)\leq
[2\alpha M+\exp(-2\alpha M)-\alpha M^{2}]\ln2
+\epsilon^{2}\alpha\ln 2\ \ \forall \ \ \epsilon > 0\ .
$$
We will show that the integrand vanishes for any given value of $t$.
Consider the first term. In this case one has 
$$
(m-M)^{2}\leq (1+M)^{2}
$$
and thus
\begin{equation}
\label{primopezzo}
\mathbb{E}\Omega_{t}[(m-M)^{2}\chi_{m\leq M-\epsilon}] \leq
(1+M)^{2}\mathbb{E}\Omega_{t}[\chi_{m\leq M-\epsilon}] \ .
\end{equation}
Similarly
\begin{equation}
\label{terzopezzo}
\mathbb{E}\Omega_{t}[(m-M)^{2}\chi_{m\geq M+\epsilon}] \leq
(1-M)^{2}\mathbb{E}\Omega_{t}[\chi_{m\geq M+\epsilon}]
\end{equation}
for the third term.
Let us proceed with the first case by dividing the interval
$[-1, M-\epsilon]$ into $L$ small sub-intervals 
$[m_{a},m_{a+1}]$, labeled by $a=1,\ldots ,L$.
We assume $m_{1}=-1, m_{L+1}=M-\epsilon$.
We can write
$$
\mathbb{E}\Omega_{t}[(m-M)^{2}\chi_{m\leq M-\epsilon}]=
\mathbb{E}\frac{Z^{\chi}_{N,t}}{Z_{N,t}}
$$
if we define
\begin{eqnarray*}
Z_{N,t} & = & \sum_{\sigma}\prod_{ij}^{1,N}
(1+\sigma_{i}\sigma_{j})^{K^{ij}_{1}}\prod_{i=1}^{N}(1+\sigma_{i})^{K^{i}_{0}}\ , \\
Z^{\chi}_{N,t} & = &  \sum_{\sigma}\prod_{ij}^{1,N}
(1+\sigma_{i}\sigma_{j})^{K^{ij}_{1}}\prod_{i=1}^{N}(1+\sigma_{i})^{K^{i}_{0}}
\chi_{m_{a}\leq m\leq m_{a+1}}\ ,
\end{eqnarray*}
assuming $m_{a}\leq m\leq m_{a+1}$ and $M$ is chosen to satisfy (\ref{extremal}).
We know from the sum rule that
$$
\lim_{N\to\infty}A_{N}(\alpha,\beta)\geq
[2\alpha M+\exp(-2\alpha M)-\alpha M^{2}]\ln2\ .
$$
If we knew that, choosing $L$ sufficiently large, we also have
$$
\frac1N\mathbb{E}\ln Z_{N,t}^{\chi}<[2\alpha M+\exp(-2\alpha M)-\alpha M^{2}]\ln 2
$$
then we would be sure that
$$
\lim_{N\to\infty}\mathbb{E}\frac{Z^{\chi}_{N,t}}{Z_{N,t}}=0
$$
because of the almost certain convergence. This would mean
that the right hand side of (\ref{primopezzo}) vanishes,
and analogously for $(\ref{terzopezzo})$. A simple interpolation
argument brings
$$
\frac1N\mathbb{E}\ln Z_{N,t}^{\chi}\leq\frac1N\mathbb{E}\ln Z_{N,0}^{\chi}
-t\alpha M^{2}\ln 2+t\alpha (M-m_{a})^{2}\ln 2\ ,
$$
since $\Omega^{\chi}_{t}[(m-M)^{2}]\leq(M-m_{a})^{2}$
if $m_{a}\leq m\leq m_{a+1}<M$, with an obvious meaning
of $\Omega^{\chi}_{t}$. 
Now define
$$
\rho(\mu)=(\exp(-2\alpha M)+2\alpha M)\ln 2+(1-M)(\ln\cosh\lambda
-\lambda\tanh\lambda)
$$
with
$$
\tanh\lambda=\left\{
\begin{array}{ll}
     \frac{\mu-M}{1-M}\ , & \mbox{if}\ \  M\leq \mu \leq 1\ ,  \\
     \frac{M-\mu}{1-M}\ , & \mbox{if}\ \  2M-1\leq \mu\leq M  \ .
\end{array}\right.
$$
Notice that $\rho(\mu)\to-\infty$ if $\mu<2M-1$, and $\tanh\lambda=0$
if $\mu=M$ and thus $\rho(M)=(\exp(-2\alpha M)+2\alpha M)\ln 2$.
We need at this point
\begin{lemma}
Given an interval $[b_{1},b_{2}]$ we have
$$
\lim_{N\to\infty}\frac1N\mathbb{E}\ln\sum_{\sigma}\prod_{i=1}^{N}
(1+\sigma_{i})^{K^{i}}\chi_{b_{1}\leq m\leq b_{2}}=\inf_{b_{1}\leq \mu\leq b_{2}}
\rho(\mu)\ .
$$
\end{lemma}
The necessity to introduce the function $\rho$ and the proof of the lemma
are the result of a standard micro-canonical analysis.
\textbf{Proof}.
Notice that $\rho$ is symmetrical in the interval $[-2M+1,1]$ with respect
to the central point $\mu=M$, and it is increasing for $2M-1\leq \mu\leq M$,
decreasing for $M\leq \mu\leq 1$. We are interested in computing
$$
\lim_{N\to\infty}\frac1N \mathbb{E}\ln\sum_{\sigma}\prod_{i=1}^{N}
[(1+\sigma_{i})^{K^{i}}\exp(\lambda\sigma_{i})]=
\mathbb{E}\ln\sum_{\sigma_{1}}
[(1+\sigma_{1})^{K^{1}}\exp(\lambda\sigma_{1})]\ .
$$
This is easy to do as $\sum_{\sigma_{1}}(1
+\sigma_{1})^{K^{1}}\exp(\lambda\sigma_{1})$
is equal to $2\cosh\lambda$ if $K^{1}=0$, to $2^{K^{1}}\exp\lambda$
if $K^{1}>0$. Hence
\begin{multline*}
\mathbb{E}\ln\sum_{\sigma_{1}}
[(1+\sigma_{1})^{K^{1}}\exp(\lambda\sigma_{1})]=\\
p_{0}(\ln 2+\ln\cosh\lambda)+\sum_{k=1}^{\infty}p_{k}(k\ln 2+\lambda)
=\\
p_{0}(\ln 2+\ln\cosh\lambda)+2\alpha M\ln 2+\lambda(1-p_{0})
\end{multline*}
where clearly $p_{0}=\exp(-2\alpha M)$ and $\sum_{k>0}p_{k}k=2\alpha M$.
Let us now compute
\begin{multline*}
\frac1N \mathbb{E}\ln\sum_{\sigma}\prod_{i=1}^{N}
(1+\sigma_{i})^{K^{i}}\chi_{m\leq\mu}\leq\\
\frac1N \mathbb{E}\ln\sum_{\sigma}\prod_{i=1}^{N}
[(1+\sigma_{i})^{K^{i}}\exp(\lambda\sigma_{i})]\exp(\lambda N(\mu-m))=\\
\lambda\mu+\exp(-2\alpha M)(\ln 2+\ln\cosh\lambda)+2\alpha M\ln 2
-\lambda(1-\exp(-2\alpha M))
\end{multline*}
since $\chi_{m\leq\mu}\leq\exp(\lambda N(\mu-m))$ for all $\lambda\geq 0$.
But now it is easy to find the minimum with respect to $\lambda$ of this
expression, which is precisely the convex 
function $\rho$ previously defined: the condition is $\tanh\lambda=(M-\mu)/(1-M)$,
as long as $2M-1<\mu\leq M$, otherwise the $\rho$ is decreasing
and the infimum is for $\lambda\to\infty$.
The case $\mu\geq M$ is analogous. This proves the lemma. $\Box$ \\
Actually,
standard micro-canonical approach would allow to prove the equality in the
statement of the lemma.
The lemma just proven implies 
$$
\frac1N\mathbb{E}\ln Z_{N,t}^{\chi}\leq \rho(m_{a+1})+t\alpha(M-m_{a})^{2}\ln 2
-t\alpha M^{2}\ln 2
$$
as $\chi_{m_{a}\leq m\leq m_{a+1}}\leq\chi_{m\leq m_{a+1}}$ trivially. We can rewrite
the previous inequality as
\begin{multline*}
\frac1N\mathbb{E}\ln Z_{N,t}^{\chi}\leq \\ 
\rho(m_{a+1})+\alpha\ln 2(M-m_{a})^{2}t
+\alpha\ln 2[(M-m_{a})^{2}-(M-m_{a+1})^{2}]t-\alpha \ln 2 M^{2} t\\
\leq (\exp(-2\alpha M)+2\alpha M-\alpha M^{2}t)+(1-M)(\ln\cosh\lambda
-\lambda\tanh\lambda)+\\
\alpha\ln 2(M-m_{a+1})^{2}+2(m_{a+1}-m_{a})
\end{multline*}
since $(M-m_{a})^{2}-(M-m_{a+1})^{2}\leq 2(m_{a+1}-m_{a})$.
As $\tanh\lambda=(M-m_{a+1})/(1-M)$, if we could now prove 
\begin{multline}\label{negativo}
(1-M)(\ln\cosh\lambda-\lambda\tanh\lambda)+\alpha\ln 2(M-m_{a+1})^{2}\equiv\\
(1-M)[\ln\cosh\lambda-\lambda\tanh\lambda+\alpha\ln 2(1-M)\tanh^{2}\lambda]
\end{multline}
to be strictly negative for all $m_{a+1}\leq M-\epsilon$, we would be done,
for it would suffice to take the partition $\{m_{a}\}$ fine enough
in order to have
$$
\lim_{N\to\infty}
\frac1N\mathbb{E}\ln Z_{N,t}^{\chi}<\exp(-2\alpha M)+2\alpha M-\alpha M^{2}t
$$
which proves the theorem.
So let us consider the function
$$
g(\lambda)=\ln\cosh\lambda-\lambda\ln 2(1-M)\tanh^{2}\lambda\ .
$$
Such a function is easily proven to be
concave decreasing in $\lambda^{2}$, and this follows from the fact that
if $M(\alpha)$ fulfills (\ref{extremal}), then it is increasing in $\alpha$.
The explicit form of the derivative of $M(\alpha)$ and these last observations
give
$$
\left.\frac{dg(\lambda)}{d\lambda^{2}}\right|_{\lambda=0}\leq -\frac12 (1-\ln 2)
$$
and from the concavity
$$
g(\lambda)\leq -\frac12(1-\ln 2)\lambda^{2}\leq -\frac12(1-\ln 2)\tanh^{2}\lambda\ .
$$
Recall that $\tanh\lambda=(M-m_{a+1})/(1-M)$, and we can now state that 
the quantity in (\ref{negativo}) is strictly negative whenever 
$m_{a+1}\leq M-\epsilon$. As anticipated, this proves the theorem.
In fact, the case $m\geq M+\epsilon$ is totally analogous.  $\Box$

\subsection{Entropy}

Define the entropy per spin of the model as
$$
s_{N}(\alpha,\beta)=A_{N}(\alpha,\beta)-\beta\partial_{\beta}A_{N}(\alpha,\beta)\ .
$$
Its value at temperature zero is
$$
s^{(0)}_{N}(\alpha)=\lim_{\beta\to\infty}s_{N}(\alpha,\beta)\ .
$$
Recall that, according to our definitions
$$
A_{N}(\alpha,\beta)=\tilde{A}_{N}(\alpha,\beta)+\ln 2+\alpha\ln\cosh\beta
$$
and notice
$$
\partial_{\beta}\tilde{A}_{N}(\alpha,\beta)=\alpha(1-\tanh^{2}\beta)
\mathbb{E}\frac{\Omega(\sigma_{i_{o}}\sigma_{j_{0}})}{1
+\Omega(\sigma_{i_{0}}\sigma_{j_{0}})\tanh\beta}\leq \frac12
\alpha(1-\tanh^{2}\beta)\ .
$$
Since $\lim_{\beta\to\infty}\beta(1-\tanh^{2}\beta)=0$, and
moreover $\lim_{\beta\to\infty}(\ln\cosh\beta-\beta\tanh\beta)=-\ln 2$,
the inequality above means that 
$$
s^{(0)}_{N}(\alpha)=\tilde{A}_{N}^{(0)}(\alpha)+(1-\alpha)\ln 2\ .
$$
We also know that
\begin{lemma}
The function $Ns_{N}^{(0)}(\alpha)$ of the size $N$ of the system
is sub-additive.
\end{lemma}
This is a consequence of Lemma \ref{existence},
and guarantees that
$\lim_{N\to\infty}s_{N}^{(0)}(\alpha)=\inf_{N}s_{N}^{(0)}(\alpha)\equiv
s^{(0)}(\alpha)$. 

Now, in the previous subsection we computed $\tilde{A}_{N}^{(0)}(\alpha)$
in the thermodynamic limit, and thus we also gained the next
\begin{theorem} The following formula 
$$
s^{(0)}(\alpha)=\sup_{M}\{-\alpha(1-M)^{2}+\exp(-2\alpha M)\}\ln 2
$$
provides the entropy per spin of the model at temperature zero,
both in the Poisson and in the Bernoulli cases.
\end{theorem}
As expected this formula prescribes the correct critical value
$\alpha=1/2$, below which the entropy decreases linearly
from $\ln 2$ to half this value: $s^{(0)}(\alpha)=(1-\alpha)\ln 2$.
Notice that $s^{(0)}(\alpha)>0$
for all values of $\alpha$ and tends to zero as $\alpha\to\infty$.

Somewhat surprisingly, the same low connectivity behavior
with strictly positive entropy is captured by the annealed 
approximation as well (see the appendix).

As a last remark, we already noticed that the free energy 
of the Viana-Bray model at high temperature is the same
as the one of our model, as a consequence the
Viana-Bray model too has strictly positive entropy
at temperature zero.

\section{Existence of the transition and the critical line}

Consider again
\begin{equation*}
\label{}
\tilde{A}_{N}(\alpha,\beta)=A_{N}(\alpha,\beta)-\ln 2-\alpha\ln\cosh\beta\ .
\end{equation*}
We saw already that in the thermodynamic limit
$\tilde{A}_{N}(\alpha,\beta)\to 0$ if the temperature is such that
$2\alpha\tanh\beta\leq 1$. We  want to show here that
this is precisely the high temperature region where the magnetization
is identically zero and that in the complementary low temperature
region the symmetric solution described above does not hold. In fact the model
exhibits a transition and the magnetization fluctuates in the 
low temperature region. More precisely we want to 
prove the following
\begin{theorem} In the low temperature region defined by
$$
2\alpha\tanh\beta>1
$$
the limiting free energy differs from the symmetric expression, i.e.
$$
\liminf_{N\to\infty}\tilde{A}_{N}(\alpha,\beta)>0\ .
$$
\end{theorem}
\textbf{Proof}. 
Let us keep $2\alpha\tanh\beta=\beta^{\prime}$ constant, so that
$$
2(\tanh\beta) d\alpha+2\alpha(1-\tanh^{2}\beta)d\beta=0\ .
$$
We have
\begin{eqnarray}
d\tilde{A}_{N}(\alpha,\beta) & = & \partial_{\alpha}\tilde{A}_{N}(\alpha,\beta)
d\alpha+\partial_{\beta}\tilde{A}_{N}(\alpha,\beta)d\beta \nonumber\\
{} & = & \mathbb{E}\ln(1+\Omega(\sigma_{i_{0}}\sigma_{j_{0}})\tanh\beta)
d\alpha \label{derivatatotale}\\
{} & {} & \quad +\alpha(1-\tanh^{2}\beta)\mathbb{E}
\frac{\Omega(\sigma_{i_{0}}\sigma_{j_{0}})}{1+
\Omega(\sigma_{i_{0}}\sigma_{j_{0}})\tanh\beta}d\beta \nonumber\\
{} & = & \left[\mathbb{E}\ln(1+\Omega(\sigma_{i_{0}}\sigma_{j_{0}})\tanh\beta)
-\mathbb{E}\frac{\Omega(\sigma_{i_{0}}\sigma_{j_{0}})\tanh\beta}{1
+\Omega(\sigma_{i_{0}}\sigma_{j_{0}})\tanh\beta}\right]d\alpha\nonumber
\end{eqnarray}
which is non-negative as $\ln(1+x)\geq x/(1+x)$ for $x\geq 0$,
and we are considering the case $d\alpha\geq 0$.
The function $\ln(1+x)-x/(1+x)$ is positive, increasing and convex 
for $0 \leq x \leq 1$. This allows for the computation of 
$\tilde{A}_{N}(\alpha,\beta)$, at least for $N\to\infty$ and $\beta\to 0$.

As anticipated, our strategy for the proof consists in considering
the limit as $\beta\to\infty$, so recall the notation
$$
\tilde{A}_{N}^{(0)}(\alpha)=
\lim_{\beta\to\infty}\liminf_{N\to\infty}\tilde{A}_{N}(\alpha,\beta)=\frac1N\mathbb{E}\ln\sum_{\sigma}
\prod_{i,j}^{1,N}(1+\sigma_{i}\sigma_{j})^{K^{ij}}-\ln 2 
$$
where once again $K^{ij}$ are independent Poisson random variables
with mean $\alpha/N$, and we know that
$$
\lim_{N\to\infty}\tilde{A}^{(0)}_{N}(\alpha)= 0 \ \ \mbox{for}\ \ \alpha\leq\frac12\ .
$$
Lemma \ref{existence}
implies $\lim_{N\to\infty}\tilde{A}_{N}^{(0)}(\alpha)=
\inf_{N}\tilde{A}_{N}^{(0)}(\alpha)\equiv\tilde{A}^{(0)}(\alpha)$,
and Theorem \ref{zeropressure} gives
$$
\tilde{A}^{(0)}(\alpha)=\max_{M}\{ 2M\alpha+\exp(-2\alpha M)
-\alpha M^{2}\}\ln 2\ .
$$
Notice that 
$$
\frac{d\tilde{A}^{(0)}(\alpha)}{d\alpha}=M^{2}(\alpha)
$$
which means $\tilde{A}^{(0)}(\alpha)>0$ for $\alpha>1/2$,
and finally implies the statement of the theorem we wanted to prove
because of (\ref{derivatatotale}). $\Box$
\begin{remark}
In this last lemma the maximum is clearly where and only where
$1-M(\alpha)=\exp(-2\alpha M(\alpha))$, which determines $\alpha=1/2$
as the critical point. In fact, $M(\alpha)=0$ for $\alpha\leq 1/2$, while
$M(\alpha)>0$ for $\alpha>1/2$. Furthermore, the critical index is given by
$M(\alpha)=2\sqrt{\alpha-1/2}+O(M^{2})$ for $\alpha>1/2$.
\end{remark}

Notice that the annealed model has the same high temperature
free energy and the same critical line as the quenched one
(see the appendix).

The interested reader might want to compare our proof with
the results obtained through replica techniques in \cite{os1}.


\section{Self-averaging}

In this section we briefly discuss the limiting self-averaging 
of the free energy density, of the magnetization, 
and we exhibit the relations coming from the
self-averaging of the internal energy. The self-averaging
properties hold for both the quenched and the annealed models
(in the latter the relations simplify as the squared magnetization
replaces the product of two randomly chosen spins).

\subsection{Magnetization}\label{self-av}

The magnetization turns out to be self-averaging, and therefore
it is an actual order parameter, although it is not clear whether 
it is the only order parameter. Let us prove the next
\begin{theorem}
The following identity 
$$ 
\mathbb{E} \Omega(m^{4}) = \mathbb{E} \Omega^{2}(m^{2})
$$
holds in the thermodynamic limit.
\end{theorem}
\textbf{Proof}. Consider the derivative with respect to the connectivity
of the mean squared magnetization. Using (\ref{main2}) we have
$$
\frac{d}{d\alpha}\mathbb{E}\Omega(m^{2})=N\mathbb{E}
[\Omega_{K+1}(m^{2})-\Omega_{K}(m^{2})]
$$
which is bounded. Proceeding further
\begin{eqnarray*}
\frac{d}{d\alpha}\mathbb{E}\Omega(m^{2}) & = & N\mathbb{E}
\frac{\Omega(m^{2})+\Omega(m^{2}\sigma_{i_{0}}\sigma_{j_{0}})
\tanh\beta}{1+\Omega(\sigma_{i_{0}}\sigma_{j_{0}})\tanh\beta}-
\mathbb{E}\Omega(m^{2}) \\
{} & = & N(\tanh\beta)\mathbb{E}\frac{\Omega(m^{2}\sigma_{i_{0}}\sigma_{j_{0}})-
\Omega(m^{2})\Omega(\sigma_{i_{0}}\sigma_{j_{0}})}{(1+{\tilde{A}^{(0)}(\alpha)}
\Omega(\sigma_{i_{0}}\sigma_{j_{0}})
\tanh\beta)(1+\Omega(m^{2})\tanh\beta)}\\
{} & {} & +N(\tanh^{2}\beta) \mathbb{E}
\frac{\Omega(m^{2})[\Omega(m^{2}\sigma_{i_{0}}\sigma_{j_{0}})-
\Omega(m^{2})\Omega(\sigma_{i_{0}}\sigma_{j_{0}})]}{(1+\Omega(\sigma_{i_{0}}\sigma_{j_{0}})
\tanh\beta)(1+\Omega(m^{2})\tanh\beta)}
\end{eqnarray*}
but this means
$$
\mathbb{E}[\Omega(m^{2}\sigma_{i_{0}}\sigma_{j_{0}})-
\Omega(m^{2})\Omega(\sigma_{i_{0}}\sigma_{j_{0}})]
=\mathbb{E} \Omega(m^{4}) - \mathbb{E} \Omega^{2}(m^{2})
\to 0
$$
when $N\to\infty$. $\Box$

Another even stronger self-averaging identity is formulated in the next
\begin{theorem} The following identity
$$
\langle m^{4}\rangle = \langle m^{2} \rangle^{2}
$$
holds in the thermodynamic limit.
\end{theorem}
We are not going to prove this here, we will show some similar relations
later in this section.

\subsection{Free energy}

As expected on a physical ground,
the free energy of our model does not
fluctuate in the thermodynamic limit. 
This is formalized in the next theorem.
\begin{theorem}
For all values of $N$, $\alpha$, and $\beta$,
the following exponential inequality
\begin{equation*}
\label{}
\mathbb{P}\left\{\left|\frac{1}{\beta N}\ln Z_{N}
-\frac{1}{\beta N}\mathbb{E}\ln Z_{N}\right|\geq\epsilon\right\}
\leq 2\exp[N(\epsilon-\alpha(1+\epsilon/\alpha)\ln(1+\epsilon/\alpha))]
\end{equation*}
estimates the probability for the random free energy
to deviate from its expectation.
\end{theorem}
\textbf{Sketched proof}. The proof 
is quite standard and
proceeds along the same lines dictated for instance in
\cite{talabook} for disordered systems. We will only sketch 
the proof here.
For a given real number $\lambda$, let us define 
$$
\Phi^{(\lambda)}_{N}(t)=\ln \mathbb{E}_{1}\exp[\lambda\mathbb{E}_{2}
\ln Z_{N}(t)]
$$
with
$$
Z_{N}(t)=\sum_{\sigma}\exp\bigg(\beta\sum_{\nu=1}^{K_{1}}
\sigma_{i_{\nu}}\sigma_{j_{\nu}}+\beta\sum_{\nu=1}^{K_{2}}
\sigma_{k_{\nu}}\sigma_{l_{\nu}}\bigg)
$$
where $K_{1}$ and $K_{2}$ are Poisson random variables with
mean $t\alpha N$ and $(1-t)\alpha N$ respectively, while
$\mathbb{E}_{1}$ and $\mathbb{E}_{2}$ denote the expectation with respect to
$\{K_{1},\{i_{\nu}\},\{j_{\nu}\}\}$ and to $\{K_{2},\{k_{\nu}\},\{l_{\nu}\}\}$ respectively.
The reason for such a construction is 
$$
\exp[\Phi_{N}^{\lambda}(1)-\Phi_{N}^{(\lambda)}(0)]=
\mathbb{E}\exp[\lambda(\ln Z_{N}-\mathbb{E}\ln Z_{N})]\ .
$$
The derivative with respect to $t$ reads
\begin{multline*}
\frac{1}{\alpha}\frac{d}{dt}\Phi^{(\lambda)}_{N}(t)=\\
\sum_{i,j}^{1,N}
\frac{\mathbb{E}_{1}\exp[\lambda\mathbb{E}_{2}\ln Z_{N}(t)
(\exp(\lambda\mathbb{E}_{2}\ln\Omega(\sigma_{i}\sigma_{j})-1)
-\lambda\mathbb{E}_{2}\ln\Omega(\sigma_{i}\sigma_{j}))]}{\mathbb{E}_{1}
\exp[\lambda\mathbb{E}_{2}\ln Z_{N}(t)]}
\end{multline*}
The simple observation that 
$$
-\beta\leq\mathbb{E}_{2}\ln\Omega[\exp(\beta\sigma_{i}\sigma_{j})]\leq\beta
$$
ensures
$$
\left|\frac{d}{dt}\Phi^{(\lambda)}_{N}(t)\right|\leq
\alpha N[\exp(|\lambda|\beta)-1-|\lambda|\beta]
$$
The validity of this bound for all $\lambda$ together with
Tchebyshev's inequality implies that the random free energy deviates
from its expectations with a probability exponentially small in the size of the system.
$\Box$

\subsection{Internal energy and self-averaging relations}

The self-averaging of the free energy implies, by standard convexity
arguments \cite{g2}, the self-averaging of the internal energy.
More explicitly, let us state without proof the next
\begin{theorem} In the thermodynamic
limit the internal energy does not fluctuate
$$
\lim_{N\to\infty}\left[\frac{\langle H^{2}_{N} \rangle}{N}-
\left(\frac{\langle H_{N} \rangle}{N}\right)^{2}\right]=0
$$
with the possible exception of a zero measure
set of values of the inverse temperature $\beta$.
\end{theorem}
From the self-averaging of the internal energy one can obtain
useful information of probabilistic nature about the thermodynamic
behavior of the model.
Since we already have an expression for the averaged internal energy,
given by (\ref{betaderivative}), it is not difficult to perform calculations
similar to all the others done so far and obtain
\begin{proposition} In the thermodynamic limit, the following identity holds
\begin{multline*}
\mathbb{E}\left[\frac{\Omega(\sigma_{i}\sigma_{j}\sigma_{k}\sigma_{l})
+\Omega(\sigma_{i}\sigma_{j})\tanh\beta
+\Omega(\sigma_{k}\sigma_{l})\tanh\beta+\tanh^{2}\beta}{1
+\Omega(\sigma_{i}\sigma_{j})\tanh\beta
+\Omega(\sigma_{k}\sigma_{l})\tanh\beta
+\Omega(\sigma_{i}\sigma_{j}\sigma_{k}\sigma_{l})\tanh^{2}\beta}\right]\\
=\left[\mathbb{E}\frac{\tanh\beta+\Omega(\sigma_{i}\sigma_{j})}{1
+\Omega(\sigma_{k}\sigma_{l})\tanh\beta}\right]^{2}
\end{multline*}
for all $i,j,k,l\in\{1,\ldots , N\}$.
\end{proposition}
This identity provides a further restriction beyond the self-averaging 
of the magnetization, characterizing the thermodynamics of the model.


\section{Outlook}

We showed that dilute mean field ferromagnets are not so poorer
than spin glasses, but not as difficult either, and this makes them
quite interesting. The annealed model is already rather
interesting, contrarily to other annealed models. 
It has a phase transition, and a non-negative zero-temperature entropy. 
The annealed model actually enjoys a strictly positive entropy at temperature zero. 
It provides the same ground state as the quenched model, and the same 
high temperature regime, with the same critical point.
The quenched model is highly non-trivial.
The control of the zero-temperature regime
we gained suggests an approach
to the model at a generic temperature, since the
kind of ``replica symmetric'' method employed here 
allows for an extension to a more general
distribution of the order parameter $\Omega(\sigma_{i}\sigma_{j})$. 
The approach based on cavity fields developed instead 
in the physical literature for spin glasses, though not fully rigorous,
also can be used to study the generic behavior. We will report
our results on these topics elsewhere \cite{dsg2}, along with 
other interesting developments: a (G)REM-like approach, the p-spin case, 
the form of the cavity fields and its consequences.
Both the annealed and the quenched models enjoys the 
stability properties known for spin glasses: a suitable cavity field 
is equivalent to the addition of one particle to the system, and this makes 
all overlaps squared. As a consequence, the free energy is the difference 
between two terms, and does not depend on certain overlap monomials, 
and this makes it possible to compute the critical exponents. This is all 
quite easy in the annealed model, while in the quenched model we have no 
proof, through our techniques at least,
of the existence of the thermodynamic limit of the free energy density 
as yet, and the procedure to compute the critical exponents in slightly more 
involved anyway. We will report on this in a separate paper \cite{dsg2}.
As a last remark, some generalizations to a bipartite model with two
populations of spins are possible \cite{dsg2}.


\appendix

\section{Annealed model}

In this appendix, we have to consider separately the expectation with respect to 
the Poisson variables and the expectation with respect to the random choice
of the spins. The expectation with respect to the random choice of the spins
is denoted by $\mathbb{E}_{s}$, while the expectation with respect
to Poisson random variables is denoted by $\mathbb{E}_{P}$.
Collectively we will use again $\mathbb{E}=\mathbb{E}_{P}\mathbb{E}_{s}$.
The annealed pressure is defined by
$$
\bar{A}_{N}(\alpha,\beta)=\frac1N\mathbb{E}_{P}\ln\mathbb{E}_{s}
\sum_{\sigma}\exp(-\beta H_{N}(\sigma))\ .
$$
We may easily include the action of an external field
on the system, by adding a term $-hNm$ to the Hamiltonian.
The next two subsections are devoted to the calculation
of the limiting pressure by means of two opposite bounds.
Both bounds are obtained following the ideas of \cite{g7,g8},
which allow for a generalization when
a quadratic function is replaced by a generic convex one.

It is easy to see that the pressure can be written as
$$
\bar{A}_{N}(\alpha,\beta)=\alpha\ln\cosh\beta+
\frac1N\mathbb{E}_{P}\ln\sum_{\sigma}
\exp(K\ln(1+m^{2}\tanh\beta))
+N\beta h m).
$$
It is thus convenient to define a function $f$ of the magnetization $m$ by
\begin{equation}
\label{effe}
f(m)=\ln(1+m^{2}\tanh\beta)\ ,
\end{equation}
whose derivative is 
\begin{equation}
\label{effeprimo}
f^{\prime}(m)=2\tanh\beta\frac{m}{1+m^{2}\tanh\beta}
\end{equation}
and its expression will be used often in the rest.

\subsection{Lower bound for the annealed pressure}

Notice that the function $f$ defined in 
(\ref{effe}) is easily verified to be convex, and therefore
$$
f(m)\geq f(M)+f^{\prime}(M)(m-M)
$$
for any given $M$. This means that we can proceed like
in \cite{g7,g8}, where the function was $f(m)=m^{2}$, and prove
the next
\begin{lemma} The following bound
\begin{multline*}
\bar{A}_{N}(\alpha,\beta)\geq \sup_{M}\bigg\{\ln 2 + \alpha\ln\cosh\beta +
\alpha\ln(1+M^{2}\tanh\beta)\\
-(2\alpha\tanh\beta)
\frac{M^{2}}{1+M^{2}\tanh\beta}+
\ln\cosh\bigg[(2\alpha\tanh\beta)
\frac{M}{1+M^{2}\tanh\beta}+\beta h
\bigg]\bigg\}
\end{multline*}
holds for all values of the size $N$ of the system.
\end{lemma}
\textbf{Proof}. From the just mentioned convexity of $f$ it is obvious that
\begin{multline*}
\label{boundlower}
\bar{A}_{N}(\alpha,\beta) \geq \alpha\ln\cosh\beta \\
+\sup_{M}\left\{\frac1N\mathbb{E}_{P}\ln\sum_{\sigma}\exp(K
[f(M)+f^{\prime}(M)(m-M)]+\beta Nhm)
\right\}\ .
\end{multline*}
This is the lower bound. We will proceed a few steps forward
to have a more explicit expression of the annealed pressure.
Let us define
\begin{eqnarray*}
A_{trial}(M) & \equiv & \alpha\ln\cosh\beta+
\frac1N\mathbb{E}_{P}\ln\sum_{\sigma}\exp (K[f(M)+f^{\prime}(M)(m-M)]
+\beta Nm)\\
{} & = &  \alpha\ln\cosh\beta+
\frac1N\mathbb{E}_{P}\ln\{\exp (K
[f(M)-f^{\prime}(M)])\\
{}&{}&\qquad\qquad\qquad \times\sum_{\sigma}\exp 
[(Kf^{\prime}(M)+\beta h
)Nm]\}\\
{} & = & \ln 2 + \alpha\ln\cosh\beta+\alpha[f(M)-f^{\prime}(M)]\\
{}&{}&\qquad\qquad\qquad +
\mathbb{E}_{P}\ln\cosh\bigg[\frac{K}{N}f^{\prime}(M)+\beta h\bigg]
\ .
\end{eqnarray*}
Using again a simple convexity argument we 
take the expectation $\mathbb{E}_{P}$ inside the $\ln\cosh$
function and obtain
$$
A_{trial}(M)\geq \ln 2 + \alpha\ln\cosh\beta
+\alpha[f(M)-f^{\prime}(M)]+
\ln\cosh[\alpha f^{\prime}(M)+\beta h]
$$
although the equal sign would hold in the thermodynamic limit
since $K/N\to\alpha$.
Using the explicit form of the function $f$ we can write
\begin{multline*}
A_{trial}(M)\geq \ln 2 + \alpha\ln\cosh\beta +
\alpha\ln(1+M^{2}\tanh\beta)\\
-(2\alpha\tanh\beta)
\frac{M^{2}}{1+M^{2}\tanh\beta}+
\ln\cosh\left[(2\alpha\tanh\beta)
\frac{M}{1+M^{2}\tanh\beta}+\beta h
\right]\ .
\end{multline*}
The final result is thus
\begin{multline*}
\bar{A}_{N}(\alpha,\beta)\geq \sup_{M}\bigg\{\ln 2 + \alpha\ln\cosh\beta +
\alpha\ln(1+M^{2}\tanh\beta)\\
-(2\alpha\tanh\beta)
\frac{M^{2}}{1+M^{2}\tanh\beta}+
\ln\cosh\bigg[(2\alpha\tanh\beta)
\frac{M}{1+M^{2}\tanh\beta}+\beta h
\bigg]\bigg\}
\end{multline*}
for any size $N$ of the system, which is precisely
the statement we wanted to prove. $\Box$

\subsection{Upper bound for the annealed pressure}

Like in \cite{g7,g8},
we will employ the following trivial identity
$$
\sum_{M}\delta_{mM}=1
$$
to prove the next
\begin{lemma}
The following bound 
\begin{multline*}
\bar{A}_{N}(\alpha,\beta)\leq \frac{\ln(2N+1)}{N}+
\sup_{M}\bigg\{\ln 2 + \alpha\ln\cosh\beta +
\alpha\ln(1+M^{2}\tanh\beta)\\
-(2\alpha\tanh\beta)
\frac{M^{2}}{1+M^{2}\tanh\beta}+
\ln\cosh\bigg[(2\alpha\tanh\beta)
\frac{M}{1+M^{2}\tanh\beta}+\beta h
\bigg]\bigg\}
\end{multline*}
holds for all value of the size $N$ of the system.
\end{lemma}
\textbf{Proof}. One has
\begin{eqnarray*}
\bar{A}_{N}(\alpha,\beta) & = & \alpha\ln\cosh\beta+\frac1N\mathbb{E}_{P}\ln
\sum_{M}\sum_{\sigma}\delta_{mM}\exp[K\ln(1+m^{2}\tanh\beta)\\
{} & {} &\hspace{7.5cm}+N\beta h m] \\
{} & = & \alpha\ln\cosh\beta+\frac1N\mathbb{E}_{P}\ln
\sum_{M}\sum_{\sigma}\delta_{mM}\exp[K f(m)]
+N\beta h m] \ ,
\end{eqnarray*}
where $f$ is again the one defined in (\ref{effe}).
But now thanks to the delta function
$$
f(m)=f(M)+f^{\prime}(m-M)
$$
so that
\begin{multline*}
A_{N}(\alpha,\beta)=\alpha\ln\cosh\beta\\
+\frac1N\mathbb{E}_{P}\ln
\sum_{M}\sum_{\sigma}\delta_{mM}\exp[K(f(M)
+f^{\prime}(M)(m-M))]+N\beta h m]\ .
\end{multline*}
At this point we trivially observe that
$$
\delta_{mM}\leq 1
$$
and thus
\begin{multline*}
\bar{A}_{N}(\alpha,\beta)\leq\alpha\ln\cosh\beta\\
+\frac1N\mathbb{E}_{P}\ln
\sum_{M}\sum_{\sigma}\exp[K(f(M)+f^{\prime}(M)(m-M))
+N\beta h m]\ .
\end{multline*}
Observe now that $M$ can only take $2N+1$ values, therefore
\begin{multline*}
\bar{A}_{N}(\alpha,\beta)\leq \alpha\ln\cosh\beta\\+
\sup_{M}\frac1N\mathbb{E}\ln[(2N+1)\sum_{\sigma}\exp
[K(f(M)+f^{\prime}(m-M))+N\beta h m]
\end{multline*}
which means
$$
\bar{A}_{N}(\alpha,\beta)\leq\frac{\ln(2N+1)}{N}+\sup_{M}A_{trial}(M)
\ ,
$$
according to the definition of $A_{trial}(M)$ given in the previous
subsection. Therefore we have proven the lemma. $\Box$.

\subsection{The annealed pressure}

We can put together the lemmas of the previous two subsections,
and summarize the final result in the thermodynamic limit.
\begin{theorem} The limiting annealed pressure is given by the formula
\begin{multline*}
\lim_{N\to\infty}\bar{A}_{N}(\alpha,\beta)
 = \sup_{M}\bigg\{\ln 2 + \alpha\ln\cosh\beta +
\alpha\ln(1+M^{2}\tanh\beta)\\
-(2\alpha\tanh\beta)
\frac{M^{2}}{1+M^{2}\tanh\beta}+
\ln\cosh\bigg[(2\alpha\tanh\beta)
\frac{M}{1+M^{2}\tanh\beta}+\beta h
\bigg]\bigg\}
\end{multline*}
for all values of $\alpha$, $\beta$ and $h$.
\end{theorem}
As we said, this follows immediately from the lemmas proven in the
previous two subsections, which together imply
$$
\lim_{N\to\infty}\bar{A}_{N}(\alpha,\beta)=\sup_{M}A_{trial}(M)
$$
and noticing again that $K/N\to\alpha$ one has statement of the theorem.

\begin{remark} Notice that the convexity of the function $f$ allows
to prove the existence of the thermodynamic limit for the free energy
per spin 
apart from its calculation, using now standard techniques.
\end{remark}

\subsection{Symmetric region}

Let us remove the external field by taking $h=0$.

In the expression of the annealed pressure just computed, 
the supremum lies where
\begin{multline*}
\frac{d}{dM}\bigg[\alpha f(M)-\alpha f^{\prime}(M)M+
\ln\cosh\alpha f^{\prime}(M)\bigg]=\\
\alpha f^{\prime\prime}(M)[\tanh(\alpha f^{\prime}(M))-M]=0
\end{multline*}
where $f$ is once again defined in (\ref{effe}).
This implies
$$
\tanh\left[(2\alpha\tanh\beta)\frac{M}{1+M^{2}\tanh\beta}\right]=M
$$
from which we deduce the critical point discriminating 
the region where there is only the zero-magnetization solution $M=0$
from the region where there are two non-zero opposite solutions $\pm M^{*}$:
$$
2\alpha\tanh\beta=1
$$
is the condition that defines the critical line. 
So thanks to the infinite connectivity limit
one can deduce the critical point of the finite 
connectivity annealed model from
the critical point of the fully connected model.

When $M=0$ the annealed pressure is
$$
\lim_{N\to\infty}\bar{A}_{N}(\alpha,\beta)=\ln 2+\alpha\ln\cosh\beta
$$
which holds where $2\alpha\tanh(\beta)\leq 1$.
Therefore the annealed and quenched pressures coincide 
in this region.

\subsection{Positivity of the entropy at temperature zero}

Recall that
\begin{definition}
The limiting entropy is
$$
\bar{s}(\beta)=\bar{A}(\alpha,\beta)-\beta \partial_{\beta}\bar{A}(\alpha,\beta)\ 
$$ 
\end{definition}
The value $s_{0}$ the entropy takes at temperature zero is
$$
\bar{s}_{0}(\alpha)\equiv\lim_{\beta\to\infty}\bar{A}(\alpha,\beta)-\beta 
\partial_{\beta}\bar{A}(\alpha,\beta)\ .
$$
Notice that when $\beta\to\infty$ 
the condition for the supremum in the formula of the annealed pressure
is attained at
$$
M=0\ \ \mbox{if}\ \ \alpha\leq\frac12\ \ ;\
\ M=\tanh\frac{2\alpha M}{1+M^{2}}
\ \ \mbox{if}\ \ \alpha>\frac12\ .
$$
When $\alpha\leq1/2$ we know
$$
\bar{A}(\alpha,\beta)=\ln 2+\alpha\ln\cosh(\beta)\ \ \forall\ \beta
$$
and it is easy to see that 
$\ln 2\geq s_{0}(\alpha)=(1-\alpha)\ln 2\geq (\ln 2)/2\geq 0$ in this case. 
Notice the high degeneracy of the ground state: when the connectivity
is too low there are too few interactions to move the entropy away
from the value $\ln 2$ it has when the absence of interactions
makes all configurations equally probable, maximizing the entropy.

Let us assume $\alpha>1/2$.
When computing a derivative of the pressure
we do not have to differentiate it with respect to $M$,
because of the supremum condition.
Notice also that $M^{2}$ increases with $\alpha$,
and in particular $M^{2}\to 1$ as $\alpha\to\infty$.
Moreover, simple calculations yield
\begin{equation}
\label{entropy}
\bar{s}_{0}(\alpha)=(1-\alpha)\ln 2+
\alpha\ln(1+M^{2})-2\alpha\frac{M^{2}}{1+M^{2}}+
\ln\cosh \left(2\alpha\frac{M}{1+M^{2}}\right)
\end{equation}
which is even in $M$ as expected (and it is $(1-\alpha)\ln 2$ for $M=0$,
recovering the case of low connectivity). This expression for 
$\bar{s}_{0}(\alpha)$
also says that the zero-temperature entropy tends to zero
when $\alpha\to\infty$.
Now
\begin{eqnarray*}
\frac{d\bar{s}_{0}(\alpha)}{d\alpha} & = & -\ln 2+\ln(1+M^{2})-2\frac{M^{2}}{1+M^{2}}
+\frac{2M}{1+M^{2}}\tanh\frac{2\alpha M}{1+M^{2}} \\
{} & = & -\ln 2 +\ln(1+M^{2})\leq 0 
\end{eqnarray*}
Notice that such a derivative tends to zero when $\alpha\to\infty$,
since in this limit $M\to 1$, and it is increasing in $\alpha$.
In other words the zero-temperature entropy is convex in the connectivity.

So the zero-temperature entropy, as a function
of the connectivity, decreases linearly from $\ln 2$ to $(\ln 2)/2$
for $0\leq \alpha \leq  1/2$, then it becomes strictly convex,
decreases monotonically, and asymptotically decays to zero.

In particular, we proved the following
\begin{proposition}
The entropy of the infinite volume annealed model remains strictly positive
$$
\bar{s}_{0}(\alpha)=\lim_{\beta\to\infty} [\bar{A}(\alpha,\beta)
-\beta \partial_{\beta}\bar{A}(\alpha, \beta)] > 0
$$
when the temperature decreases to zero.
\end{proposition}
\begin{remark}
We encountered an annealed model
without the problem of a negative entropy: on the contrary, it has
a strictly positive entropy with a highly degenerate ground state,
and the entropy vanishes only when the number of interactions is large
enough, namely when $\alpha\to\infty$.
\end{remark}

Notice that in the low connectivity region the annealed entropy is
the same as the one of the quenched model.

\subsection{Infinite connectivity limit}

We already saw that both the quenched and the annealed model
reduce to the fully connected one in the infinite connectivity limit.
The same result can be obtained directly from  
the formula of the free energy we found (only in 
the thermodynamic limit) which reduces in the infinite connectivity limit
to the well known formula for the free energy of the Curie-Weiss model.



\section*{Acknowledgments}

The authors thank Andrea Montanari for precious comments and remarks.
LDS ackowledges partial support by the CULTAPTATION project
(European Commission contract FP6-2004-NEST-PATH-043434).


\end{document}